\documentclass[prl,aps,superscriptaddress,footinbib,twocolumn]{revtex4-2}
\usepackage[latin9]{inputenc}
\usepackage{color}
\usepackage{amsmath}
\usepackage{amssymb}
\usepackage{stmaryrd}
\usepackage{csquotes}
\usepackage{graphicx}
\usepackage[unicode=true,bookmarks=true,bookmarksnumbered=false,bookmarksopen=false,breaklinks=false,pdfborder={0 0 1},backref=false,colorlinks=true]
 {hyperref}
\hypersetup{
 linkcolor=magenta, urlcolor=blue, citecolor=blue, pdfstartview={FitH}, unicode=true}
 
\makeatletter

\usepackage{algorithm}
\usepackage{algorithmic} 
\usepackage{amsfonts}\usepackage{tabularx}\usepackage{dcolumn}\usepackage{bm}\usepackage{graphicx}\usepackage{epstopdf}
\usepackage{times}
\usepackage{braket}
\hypersetup{urlcolor=blue}

\newtheorem{definition}{Definition}

\def\ket#1{\left|#1\right\rangle}
\def\bra#1{\left\langle#1\right|}

\makeatother

\begin{document}
\title{Quantum Delegated and Federated Learning via Quantum Homomorphic Encryption}

\affiliation{Center for Quantum Information, IIIS, Tsinghua University, Beijing 100084, China\\
$^2$ Hefei National Laboratory, Hefei 230088, China\\
$^3$ Shanghai Qi Zhi Institute, Shanghai 200232, China}

\author{Weikang Li$^{1}$}
\author{Dong-Ling Deng$^{1,2,3}$}
\email{dldeng@tsinghua.edu.cn}

\begin{abstract}
Quantum learning models hold the potential to bring computational advantages over the classical realm.
As powerful quantum servers become available on the cloud, ensuring the protection of clients' private data becomes crucial.
By incorporating quantum homomorphic encryption schemes,
we present a general framework that enables quantum delegated and federated learning with a computation-theoretical data privacy guarantee.
We show that learning and inference under this framework feature substantially lower communication complexity compared with schemes based on blind quantum computing.
In addition, in the proposed quantum federated learning scenario, there is less computational burden on local quantum devices from the client side, since the server can operate on encrypted quantum data without extracting any information.
We further prove that certain quantum speedups in supervised learning carry over to private delegated learning scenarios employing quantum kernel methods.
Our results provide a valuable guide toward privacy-guaranteed quantum learning on the cloud, 
which may benefit future studies and security-related applications.
\end{abstract}

\maketitle

\emph{Introduction.}---Quantum machine learning exhibits a novel paradigm of learning and inference based on data~\cite{Biamonte2017Quantum,Dunjko2018Machine,DasSarma2019Machine,Cerezo2022Challenges},
which is promising to bring advantages over classical methods for certain learning tasks~\cite{Rebentrost2014Quantum,Liu2021Rigorous,Huang2022Quantum,Molteni2024Exponential}.
These advantages mainly focus on the computational complexity, e.g., the running time and number of samples required to build the learning model.
To obtain such quantum-versus-classical learning advantages,
it is usually required to make hardness assumptions on certain computational problems~\cite{Liu2021Rigorous,Gyurik2023Exponential}, or utilize quantum correlations for unconditional proofs~\cite{Gao2022Enhancing,Zhang2024Quantumclassical}.
With a fully-fledged quantum computer featuring a large number of individually addressable and high-fidelity qubits,
such learning advantages could be experimentally demonstrated.
Along this line, a long-term goal is to achieve advantageous applications for practical tasks and benefit other fields~\cite{Daley2022Practical}.

However,
aside from the computational aspect,
the security issue is also crucial for near-term and future quantum applications~\cite{Liu2024Quantum,Sheng2017Distributed,Hai2024Data,Caro2024Classical}.
As this field progresses,
the early generations of publicly available quantum computers are most likely expensive and only presented in the form of cloud quantum servers.
For learning tasks, a client could upload the training data as well as the learning algorithm to the quantum server.
After the server completes the learning procedure,
the results will be sent back to the client for further use.
In this delegated learning scenario,
a natural question arises:
How can one ensure that the client's data or computation is kept private from the server?
Indeed, a malicious server may try to infer from the learning procedure or even disobey the instructions from the client to extract sensitive information.
It is therefore of both theoretical and practical importance to develop privacy-preserving delegated learning protocols.

The exploration of private delegated quantum computations dates back to an interactive protocol~\cite{ChildsSecure}, where a client, Alice, with limited quantum capabilities, delegates a task to a more powerful quantum server, Steve.
Within the delegation procedure, 
a quantum one-time pad scheme is applied to hide the information of a quantum state~\cite{Ambainis2000Private}.
The following works along this direction are mainly divided into two categories---blind quantum computing and quantum homomorphic encryption.
In blind quantum computing, the client utilizes interactive protocols to hide all information from the server, including the input, output, and computation~\cite{Broadbent2009Universal}.
This framework has been developed and applied to enhance the security of quantum learning tasks~\cite{Li2021Quantum, Li2024Blind}.
In comparison, quantum homomorphic encryption focuses on quantum operations to be performed on encrypted data in such a way that the underlying plaintext data remains hidden from the server.
During the whole delegated computation process, there is only one round of interaction between the two parties~\cite{Broadbent2015Quantum,Mahadev2020Classical,Dulek2016Quantum,Brakerski2018Quantum,Ma2022Quantum,Ouyang2018Quantum,Tham2020Experimental} and it is shown feasible to support variational quantum algorithms~\cite{Li2024Secure}.
We emphasize a key difference between the two  approaches.
In blind quantum computing, the server is treated as a ``dumb'' entity that is completely unaware of the computations performed.
In contrast, quantum homomorphic encryption allows the server to execute computations on encrypted quantum data, where the server knows the algorithm being applied without gaining any information of the processed data.

\begin{figure*}[t]
\center
\includegraphics[width=0.9\linewidth]{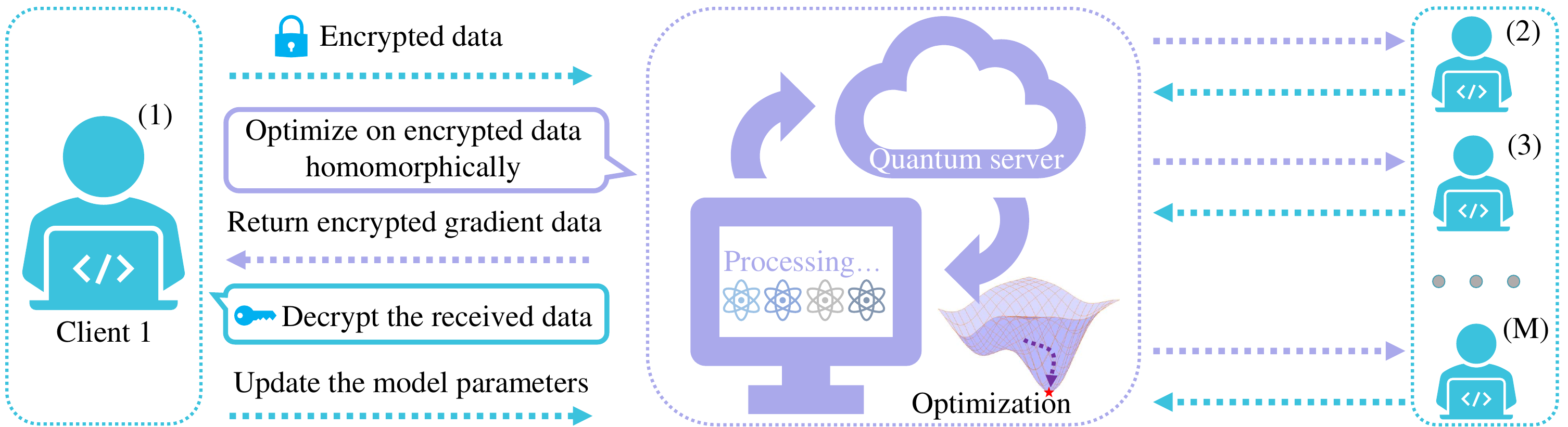}
\caption{A schematic illustration of quantum delegated and federated learning adapting quantum homomorphic encryption techniques.
On the left side, we exhibit single-client quantum delegated learning.
For the training data in the form of quantum states or classical bits,
the client applies a quantum or classical one-time pad to encrypt it, respectively.
Upon receiving the data, the server homomorphically operates on the encrypted data and returns the encrypted results, which contain the information for model optimization, to the client.
After decrypting the results, the client could then update the model parameters.
On the right side, the protocol is extended to the multi-party federated learning scenario, 
where different clients, each holding their private data, can collaboratively train a shared model.
}
\label{fig1}
\end{figure*}

In this work, we leverage ideas of quantum homomorphic encryption and present a general framework for both quantum delegated learning and federated learning, which further supports delegated inference after the learning process.
We first construct a quantum classification model and adapt it to the quantum homomorphic encryption scenario.
For the training samples in the form of general quantum states,
it is required to encrypt these states before sending them to the quantum server.
The non-trivial role of quantum homomorphic encryption is reflected by the fact that it allows the quantum server to manipulate the encrypted states, and further provide the desired outputs in an encrypted form.
After receiving the outputs from the server,
the client can efficiently recover the correct results by decryption.
This feature enables us to design delegated optimization strategies in a privacy-preserving fashion.
We further discuss several intriguing aspects of learning under this framework,
including lower communication complexity, less demand for local computational power, higher compatibility with error correction schemes, and provable quantum speedups with kernel methods.

\vspace{.1cm}
\emph{Notations and theoretical background.}---We start with the theoretical framework for quantum delegated learning with quantum homomorphic encryption.
Given a general $n$-qubit quantum state $\ket{\psi_x}$,
an information-theoretical secure encryption way is to apply the quantum one-time pad to this state:
\begin{equation}
\label{qotp}
\ket{\psi_x} \rightarrow (Z^{a_1} \otimes \dots \otimes Z^{a_n}) (X^{b_1} \otimes \dots \otimes X^{b_n}) \ket{\psi_x},
\end{equation}
where $Z$ and $X$ denote Pauli-$Z$ and Pauli-$X$ gates respectively, $a_i$ and $b_i$ are encryption keys chosen from $\{0,1\}$ randomly and independently~\cite{Ambainis2000Private}.
For anyone without the keys $a_i$ and $b_i$, this encrypted state is equivalent to a maximally mixed state and thus no information can be extracted. 
Since homomorphic encryption works with encrypted data, it would be desirable if compatible schemes could be designed for the above one-time-padded data to achieve high-level security.
Yet, this is challenging unless either (1) the client sends over certain quantum states which contain information about the keys and can be exploited to implement quantum operations on the server, or (2) the information-theoretical security, which is strong, is converted to a computation-theoretical one, assuming the hardness of certain problems, such as learning with errors~\cite{Broadbent2015Quantum,Mahadev2020Classical,Dulek2016Quantum,Brakerski2018Quantum,Ouyang2018Quantum,Tham2020Experimental,Ma2022Quantum,Fisher2014Quantum}.

Here, we adapt the schemes from Refs.~\cite{Mahadev2020Classical,Dulek2016Quantum,Brakerski2018Quantum,Ma2022Quantum} as technical subroutines for our framework.
Instead of only encrypting the quantum data,
the homomorphic encryption scheme includes two parallel computations.
Let $\hat{f}(\mathbf{a},\mathbf{b})$ denote the quantum encryption operation in Eq.~\eqref{qotp} and $\mathrm{Enc}(\mathbf{a},\mathbf{b})$ be a classical encryption function, where $\mathbf{a},\mathbf{b}$ denote the classical keys $a_1,\dots,a_n$ and $b_1,\dots,b_n$, respectively.
For a given quantum sample $\ket{\psi_x}$, the client generates classical keys $\mathbf{a},\mathbf{b}$, and send both the encrypted state $\hat{f}(\mathbf{a},\mathbf{b})\ket{\psi_x}$ and encrypted classical keys $\mathrm{Enc}(\mathbf{a},\mathbf{b})$ to the server.
By applying the quantum homomorphic encryption protocol,
the server homomorphically applies a target operation $U$ to the original state after receiving these two pieces of information.
The data processing can be illustrated as
\begin{eqnarray}
\label{encryp}
\begin{split}
\mathrm{Enc}(\mathbf{a},\mathbf{b}) \rightarrow \mathrm{Enc}(\mathbf{a}^{\prime},\mathbf{b}^{\prime}), \\
\hat{f}(\mathbf{a},\mathbf{b})|\psi_x\rangle \rightarrow \hat{f}(\mathbf{a}^{\prime},\mathbf{b}^{\prime}) U|\psi_x\rangle,
\end{split}
\end{eqnarray}
where the encryption key is updated to $\mathbf{a}^{\prime},\mathbf{b}^{\prime}$ and the server can homomorphically compute $\mathrm{Enc}(\mathbf{a}^{\prime},\mathbf{b}^{\prime})$ without access to the true values of $\mathbf{a}^{\prime}$ and $\mathbf{b}^{\prime}$~\cite{Mahadev2020Classical}.
Besides, the setting for handling general quantum states can be relaxed to states on the computational basis in certain scenarios, which allows a purely classical client and will be discussed in the section of quantum learning advantages with kernel methods.

\vspace{.1cm}
\emph{Delegated learning with an untrusted server.}---We first consider variational quantum classifiers, which is applicable to near-term quantum devices, in our delegated learning setting~\cite{Cerezo2021Variational}.
The learning model is built on a parameterized quantum circuit denoted by $U_{\boldsymbol{\theta}}$, where $\boldsymbol{\theta}$ represents the collection of trainable parameters.
For a training sample $\ket{\psi_x(i)}$ indexed by $i$, the output of this model is an expectation value $\bra{\psi_x(i)}U^{\dagger}_{\boldsymbol{\theta}}OU_{\boldsymbol{\theta}}\ket{\psi_x(i)}$ for a certain observable $O$.
By designing appropriate cost functions to measure the distance between the current output value and the target one,
an optimization procedure can be applied to capture the data pattern and update circuit parameters.
Assuming the label of this sample is $y(i)$, the mean square error can be used as a cost function:
\begin{equation}
\label{mse}
\mathcal{C}_{\mathrm{MSE}}=\frac{1}{N}\sum_i(\bra{\psi_x(i)}U^{\dagger}_{\boldsymbol{\theta}}OU_{\boldsymbol{\theta}}\ket{\psi_x(i)}-y(i))^2,
\end{equation}
where $N$ denotes the size of the training set.

The delegated learning procedure begins with encrypting the training data according to Eqs.~\eqref{encryp} on the client's side.
To compile the variational quantum learning model in the homomorphic encryption scenario,
it is worth considering the encryption protocol allowing efficient implementation of arbitrary single-qubit gates~\cite{Ma2022Quantum}.
After the server homomorphically processes the data, the output quantum state becomes $\hat{f}(\mathbf{a}^{\prime},\mathbf{b}^{\prime}) U_{\boldsymbol{\theta}}|\psi_x\rangle$ with updated classical keys.
In general, the prediction of the learned model is made according to the expectation value of certain local observables.
Without loss of generality, we choose a two-label classification task and a Pauli-$Z$ measurement on qubit indexed by $k$, i.e. $Z_k$, as the prediction.
The classification can be made by defining the expectation value of $Z_k$ over zero as class $1$, and otherwise as class $2$.
However, since the output state is encrypted by $\mathbf{a}^{\prime}$ and $\mathbf{b}^{\prime}$, we need to decode the measured values from the server accordingly.
We note that $\bra{\psi_x(i)}U^{\dagger}_{\boldsymbol{\theta}}\hat{f}(\mathbf{a}^{\prime},\mathbf{b}^{\prime})Z_k\hat{f}(\mathbf{a}^{\prime},\mathbf{b}^{\prime})U_{\boldsymbol{\theta}}\ket{\psi_x(i)}$ can be simplified to $(-1)^{b^{\prime}_k}\bra{\psi_x(i)}U^{\dagger}_{\boldsymbol{\theta}}Z_kU_{\boldsymbol{\theta}}\ket{\psi_x(i)}$, where $b^{\prime}_k$ is the $k$-th element in $\mathbf{b}^{\prime}$.
This indicates that the sign of the target output is also encrypted.
On the server's side,
only the encrypted classical keys and encrypted states are available and it is impossible to obtain the correct output.
On the contrary, the client holds the decryption key to the function $\mathrm{Enc}()$ and thus can efficiently decrypt the value of $b^{\prime}_k$ after receiving $\mathrm{Enc}(\mathbf{a}^{\prime},\mathbf{b}^{\prime})$, after which the correct output value can be deciphered from the encrypted one.

The above idea has a direct application in delegated inference on a powerful quantum server which, for example, is equipped with a large-scale and fine-tuned quantum learning model.
Such models may be built on sophisticated platforms and thus expensive, only deployed on the cloud and allowing clients to send computation queries remotely.
For a general quantum state sample,
a quantum one-time pad encrypts it first and the quantum learning model can make predictions on the encrypted data following the quantum homomorphic encryption protocol.
Assume the output obtained by the server is $w$.
With decryption keys of $\mathrm{Enc}()$, only the client can decipher the classical keys $\mathbf{a}^{\prime},\mathbf{b}^{\prime}$ and make correct classification of this sample according to $(-1)^{b^{\prime}_k}\mathrm{sign}(w)$.

In addition to delegated inference, learning in a delegated fashion is also important in data-sensitive scenarios, e.g., an institution holds private health data while a server holds high computational power.
This task reduces to implementing optimization under the homomorphic encryption scheme.
Back to Eq.~\eqref{mse}, our goal is to optimize the variational parameters $\boldsymbol{\theta}$ to minimize the cost function over the training set.
During this procedure, a crucial step is to calculate the gradients of the cost function with respect to these parameters.
To this end, we can apply finite difference methods or the parameter-shift rule~\cite{Mitarai2018Quantum,Schuld2019Evaluating}, both of which boil down to measuring the expectation values of the same observable by shifting certain parameters in the quantum circuit.
Under the proposed delegated learning framework, 
the server can flexibly shift the model parameters and obtain the desired gradients in an encrypted form.
The client decrypts the gradients and uploads the updated parameters to the server, which completes an optimization iteration.

\vspace{.1cm}
\emph{Extension to federated learning.}---Federated machine learning, also known as collaborative training, allows the training of a shared model across multiple parties while keeping their private training data safe~\cite{Kairouz2021Advances}. 
Developing techniques along this line is crucial for scenarios where the data is sensitive and each party only holds a small share insufficient for training, such as limited patients' data held by different hospitals.
Federated learning is traditionally achieved by aggregating model updates computed by local parties to a central server~\cite{Kairouz2021Advances}, which has been extended to the quantum domain recently~\cite{Ren2024Quantum,Chehimi2024Foundations,Li2021Quantum,Chen2021Federated,Du2022Distributed,Zhao2023NonIID}. Yet, such a conventional approach would suffer from a trade-off between privacy preserving and local computational power requirement: 
if each party computes on their own devices and uploads the gradients to collaboratively train a model, there are high requirements for local computational power;
Instead, if they upload their data directly to a central server to release local computational burden,
a malicious server may violate their data privacy.

With the federated learning framework built upon quantum homomorphic encryption and delegated learning,
we overcome the above trade-off and achieve both data privacy and the utilization of remote computational power.
Suppose there are $M$ independent parties, each holding their dataset with a limited size.
A learning epoch proceeds by enumerating all the parties.
For each party, a subset of samples are randomly selected.
The samples are encrypted using the quantum one-time pad before being sent to the quantum server with the encrypted classical keys $\mathrm{Enc}(\mathbf{a},\mathbf{b})$.
With a similar idea in delegated learning,
by decrypting $\mathrm{Enc}(\mathbf{a}^{\prime},\mathbf{b}^{\prime})$ to obtain the updated classical keys,
the party can recover the correct gradient values of a given cost function.
After taking a weighted sum of the calculated gradients over the selected samples in this party's dataset,
an update of model parameters is then uploaded to the quantum server.
Such optimization can be iteratively applied until the training converges (Algorithm~\ref{algo_qfl}).

It is worth mentioning that, since the model parameters are publicly trained, there is inevitable private information flowing from training samples to the server.
Previous works from both quantum~\cite{Li2021Quantum,Li2024Privacypreserving} and classical~\cite{Zhu2019Deep} community have shown that the uploaded gradients can be utilized to infer the input data through reverse engineering.
This is an intrinsic feature for training a public model.
To defend against potential attacks,
various strategies can be applied as additional subroutines in the original protocol.
An effective way is to adapt differential privacy and add appropriate noise to the calculated gradients, which provide an information-theoretical bound on the leaked privacy~\cite{Abadi2016Deep,Li2021Quantum}.
In the quantum learning setting, secure inner product estimation and secure weighted gradient summation techniques have also been explored and can be utilized to enhance privacy protection~\cite{Li2024Privacypreserving}.

\begin{figure}
\begin{algorithm}[H]
\caption{Quantum federated learning}
\label{algo_qfl}
\begin{algorithmic}
\REQUIRE The model $h$ with parameters $\boldsymbol{\theta}_0$, the number of clients $M$, the number of iterations $T$
\ENSURE The trained model $h(\boldsymbol{\theta}^*)$
\STATE Initialization: Generate a length-$T$ random string $R$; each element in $R$ corresponds to an index of a client
\FOR{ $i \in [T]$}
    \STATE 1) Choose the $R_i$-th client and randomly choose a subset of training samples held by the client
    \STATE 2) Calculate the average gradients over the selected training samples according to the protocol based on quantum homomorphic encryption 
    \STATE 3) Update the model parameters according to the obtained gradients $\boldsymbol{\theta}_{i+1}\leftarrow\boldsymbol{\theta}_i$, where strategies such as differential privacy can be adapted to reduce privacy leakage in model updates
\ENDFOR
\STATE Output the trained model $h(\boldsymbol{\theta}_T)$
\end{algorithmic}
\end{algorithm}
\end{figure}

\vspace{.1cm}
\emph{Analysis.}---The framework introduced for quantum delegated and federated learning through quantum homomorphic encryption bears several intriguing merits. 
First, this approach reduces local computational burdens, ensuring that training on edge quantum devices is not necessary for data owners. 
For this point, delegated learning based on blind quantum computing also provides a solution.
The difference lies in the fact that, unlike blind quantum computing where the server is treated as an unknowing entity, the server in our framework is fully capable of performing computations and knowing the algorithm without access to the original data. 
From an information-theoretical point of view, if the client hides all the information about the data and computation from the server - as achieved in the protocols based on blind computing - then for each round of delegated learning, the client has to transfer all the information about the quantum learning model to the server which might be a huge amount even in a classical description.
In contrast, here in the proposed scheme we only focus on protecting the private data information, while the server holds the model to manipulate the encrypted data.
In this way, the client only needs to send the training samples to the server and participate in the optimization procedure, while there is no need to hold and transfer the model information.
A direct consequence of this feature is that there is substantially less communications required between the server and clients during the learning and inference processes, minimizing overhead in a distributed network.

\emph{{Error correction.}---}Aside from the communication complexity aspect, since the server directly works on the encrypted data and is in full control of executing the quantum circuit,
quantum error correction schemes can be naturally applied without involving the client~\cite{Terhal2015Quantum}.
For protocols based on blind computing subroutines,
hiding all the computational details from the server becomes a double-edged sword:
On the one hand,
perfect privacy can be achieved in single-party delegated learning.
On the other hand, however,
the client does not have detailed device information on the server, and the server has no knowledge about the delegated computation.
Thus, error correction becomes comparably more demanding, with additional overhead and efforts on both the client and server for fault-tolerant delegated quantum learning~\cite{Morimae2012Blind,Gheorghiu2015Robustness,Chien2015FaultTolerant}.

\emph{{Computational learning advantage.}---}In the delegated learning scenario, designing quantum learning tasks with computational advantages is more challenging.
On the one hand,
transferring training data to the server or preparing initial states remotely will bring additional computational costs, and quantifying quantum speedups becomes more subtle:
For example, in the well-known quantum support vector machine protocol~\cite{Rebentrost2014Quantum}, by querying a data oracle, the learning model finally achieves the running time scaling as $\mathcal{O}(\log(Nd))$, where $N$ and $d$ denote the number of samples and dimension of the feature space, respectively.
This features an exponential speedup over the classical methods.
However, in the delegated learning setting, transferring such a dataset immediately brings $\mathcal{O}(Nd)$ complexity, which nullifies all the speedups.
On the other hand,
cryptographic methods are applied to operate on encrypted data,
which brings additional overheads on the server's side.
To exhibit speedups in quantum learning under our framework,
we consider the case of learning with classical data~\cite{Liu2021Rigorous,Jerbi2024Shadows,Sweke2021Quantum}.
More specifically, we consider a concept class defined based on the discrete logarithm problem~\cite{Liu2021Rigorous,Gyurik2023Exponential}:
\begin{definition}\label{def_dlp}
For an $n$-bit prime number $p$ with a generator $a$ of $\mathbb{Z}_p^*$, we define the concept class based on the discrete logarithm problem as $\mathcal{C}_n^{\mathrm{DLP}}=\left\{c_i\right\}_{i \in \mathbb{Z}_p^*}$, where
\begin{align}
c_i(x)= \begin{cases}+1, & \text { if } \log _a x \in\left[i, i+\frac{p-3}{2}\right], \\ -1, & \text { otherwise. }\end{cases}
\end{align}
\end{definition}
Assuming the hardness of the discrete logarithm problem, it is classically intractable to learn the above concept class \cite{Liu2021Rigorous}.
For a quantum learner, for any sample $x$, it can efficiently prepare the feature state $|\phi(x)\rangle=\frac{1}{\sqrt{2^k}} \sum_{i=0}^{2^k-1}\left|x a^i\right\rangle$ where $k=n-t\log n$ for a constant $t$~\cite{Liu2021Rigorous}.
By estimating the inner products of these feature states,
we obtain a kernel matrix that further allows a classical computer to build a linear classifier.

In our proposed framework,
since the data is in the form of bit strings,
the encrypted state $\hat{f}(\mathbf{a},\mathbf{b})|\psi_x\rangle$ is also in the computational basis.
Thus, we can instead send the corresponding encrypted classical bit string to the server and let the server prepare the initial state, releasing the requirement of limited quantum power for the clients. In other words, the clients can be purely classical without any quantum power in our scenario.
By sending both the data and ancillary bits in encrypted forms and homomorphically applying the feature mappings,
the server obtains the encrypted feature state $\frac{1}{\sqrt{2^k}} \sum_{i=0}^{2^k-1}\hat{f}(\mathbf{a},\mathbf{b})\left|x a^i\right\rangle$.
To estimate the kernel element between two samples $x_1$ and $x_2$,
we apply the same quantum one-time pad $\hat{f}(\mathbf{a},\mathbf{b})$ so that the inner product of the two encrypted output states is exactly the desired kernel element.
During this process, sending classical bit strings takes $\mathcal{O}(N^2d/\epsilon^2)$ additional complexity with $N$, $d$, and $\epsilon$ being the number of samples, the data dimensionality, and the target additive error, respectively. 
Meanwhile, quantum homomorphic encryption brings a constant overhead for each gate.
Thus in all, the delegated learning framework takes at most polynomial computational overheads, and the exponential quantum-classical learning separation still holds.
We mention that the above computational learning advantage is not limited to the discrete logarithm concept class.
By designing proper learning problems, similar protocols may be implemented, e.g., for the discrete cube root problems~\cite{Kearns1994Introduction,Gyurik2023Exponential},
with a final goal towards practical and real-life applications.

\vspace{.1cm}
\emph{Conclusions and outlooks.}---In this work, we have developed a quantum delegated and federated learning framework based on quantum homomorphic encryption, which guarantees computation-theoretic privacy while leveraging the computational power of quantum servers. 
Our framework allows for training and inference on encrypted data with reduced communication complexity, making it a promising approach for future quantum cloud services. 
The delegation of computation to quantum servers not only removes the computational burden for clients but also preserves provable quantum speedups in certain learning tasks employing kernel methods.

This work makes a primary step forward in enabling delegated, secure, scalable, and efficient quantum machine learning systems. Future studies could focus on extending the framework to more complex learning models and improving the privacy guarantees through advanced cryptographic techniques. 
In addition, integrating cutting-edge classical learning models into the framework could further enhance its real-world applicability.
While promising, several challenges remain to be addressed. One issue involves managing the inevitable information leakage that occurs when publicly trained models, such as federated quantum learning models, are reverse-engineered to infer private training data. Although techniques like differential privacy can mitigate such risks as discussed above, additional work is needed to ensure robust protection in other scenarios. 
Finally, while the current framework reduces local computational requirements to a certain degree, the quantum homomorphic encryption add a constant yet large prefactor to the overall complexity on the server's side. This renders its experimental demonstration with current noisy intermediate-scale quantum devices unattainable. Developing more efficient protocols that could further enhance scalability and minimize computational complexity would be crucial for both near-term and future applications.

\vspace{.2cm}
\noindent\textbf{Note}
All the data for this study will be organized and made publicly available for download on Zenodo/Figshare/Github upon publication.

\vspace{.2cm}
\noindent\textbf{Acknowledgement} We thank Qi Ye, Si Jiang, Junyu Liu, Dong Yuan for helpful discussions. This work is supported by the National Natural Science Foundation of China (Grants No.~12075128 and No.~T2225008), the Innovation Program for Quantum Science and
Technology (No.~2021ZD0302203), Tsinghua University Dushi Program, and the Shanghai Qi Zhi Institute.

\bibliography{QMLBib}

\end{document}